\documentclass[%
reprint,
superscriptaddress,
showpacs,preprintnumbers,
 amsmath,amssymb,
 aps,
 nofootinbib,
]{revtex4-1}

\usepackage{graphicx}

\newcommand{\fig}[1]{Fig.~\ref{#1}}
\newcommand{\tab}[1]{Table~\ref{#1}}
\newcommand{\Sec}[1]{Sec.~\ref{#1}}

\begin{document}

\title{Cross section of $^3$He($\alpha$,$\gamma$)$^7$Be around the $^7$Be proton separation threshold}

\author{T.~Sz\"ucs}%
\altaffiliation[\vspace{-6mm} Zentrum Dresden-Rossendorf (HZDR), Dresden, Germany ]{szucs.tamas@atomki.mta.hu; Present address: Helmholtz-}%
\affiliation{Institute for Nuclear Research (MTA Atomki), Debrecen, Hungary}%

\author{G.~G.~Kiss}%
\affiliation{Institute for Nuclear Research (MTA Atomki), Debrecen, Hungary}

\author{Gy.~Gy\"urky}%
\affiliation{Institute for Nuclear Research (MTA Atomki), Debrecen, Hungary}

\author{Z.~Hal\'asz}%
\affiliation{Institute for Nuclear Research (MTA Atomki), Debrecen, Hungary}

\author{T.~N.~Szegedi}%
\affiliation{Institute for Nuclear Research (MTA Atomki), Debrecen, Hungary}
\affiliation{University of Debrecen, Debrecen, Hungary}

\author{Zs.~F\"ul\"op}%
\affiliation{Institute for Nuclear Research (MTA Atomki), Debrecen, Hungary}

\begin{abstract}
\begin{description}
\item[Background] The $^3$He($\alpha$,$\gamma$)$^7$Be reaction is a widely studied nuclear reaction; however, it is still not understood with the required precision. It has a great importance both in Big Bang nucleosynthesis and in solar hydrogen burning. The low mass number of the reaction partners makes it also suitable for testing microscopic calculations.
\item[Purpose]  Despite the high number of experimental studies, none of them addresses the $^3$He($\alpha$,$\gamma$)$^7$Be reaction cross sections above 3.1-MeV center-of-mass energy. Recently, a previously unobserved resonance in the $^6$Li(p,$\gamma$)$^7$Be reaction suggested a new level in $^7$Be, which would also have an impact on the $^3$He($\alpha$,$\gamma$)$^7$Be reaction in the energy range above 4.0\,MeV. The aim of the present experiment is to measure the $^3$He($\alpha$,$\gamma$)$^7$Be reaction cross section in the energy range of the proposed level.
\item[Method] For this investigation the activation technique was used. A thin window gas-cell target confining $^3$He gas was irradiated using an $\alpha$ beam. The $^7$Be produced was implanted into the exit foil. The $^7$Be activity was determined by counting the $\gamma$ rays following its decay by a well-shielded high-purity germanium detector.
\item[Results] Reaction cross sections have been determined between $E_{cm} = 4.0 - 4.4$\,MeV with 0.04-MeV steps covering the energy range of the proposed nuclear level. One lower-energy cross-section point was also determined to be able to compare the results with previous studies.
\item[Conclusions] A constant cross section of around 10.5\,$\mu$barn was observed around the $^7$Be proton separation energy. An upper limit of 45\,neV for the strength of a $^3$He($\alpha$,$\gamma$)$^7$Be resonance is derived.
\end{description}
\end{abstract}

\maketitle

\vspace{-3mm}
\section{\label{sec:intro} Introduction}
The $^3$He($\alpha$,$\gamma$)$^7$Be reaction is important both in models of Big Bang nucleosynthesis (BBN) \cite{Coc17-IJMPE} and in the {\it pp} chain of solar hydrogen burning \cite{Serenelli13-PRD}.
It was extensively studied in recent years to gain high-precision information on the reaction rate in both scenarios.
Most of the studies concentrated on the cross section at low energy to constrain the extrapolations \cite[and references therein]{Adelberger11-RMP} to the solar energy range ($E_{cm}=15-30$\,keV). A pioneering study by the LUNA collaboration \cite{Bemmerer06-PRL, Confortola07-PRC, Gyurky07-PRC, Costantini08-NPA} provided direct high-precision cross-section data in the BBN energy range, which ruled out the $^3$He($\alpha$,$\gamma$)$^7$Be nuclear physics solution of the cosmological $^7$Li problem \cite{Fields11-ARNPS}.
Even though the solar energy range is not accessible using the current experimental techniques, the reaction rate can be predicted by extrapolation. One way of extrapolation is the \mbox{R-matrix} approach, which is based on experimental data.
The R-matrix framework was applied by Descouvemont \textit{et al.} \cite{Descouvemont04-ADNDT} to determine BBN reaction rates; however, the $^3$He($\alpha$,$\gamma$)$^7$Be reaction was treated only as purely external capture leading to a constant S-factor towards higher energies. This assumption was supported by the data of Parker and Kavanagh \cite{Parker63-PR} which was the only available experimental data set at that time extending to higher energies.
Later, a precision measurement with many data points had been performed by the ERNA collaboration ranging up to $E_{cm}=3.1$\,MeV \cite{DiLeva09-PRL}. These new data contradict the former one suggesting a higher slope of the S factor. 
This discrepancy ignited a few new studies, which all support the positive slope of the S factor \cite{Carmona-Gallardo12-PRC, Bordeanu13-NPA, Kontos13-PRC}, but there are no experimental cross-section data above $E_{cm}=3.1$\,MeV. 
In the most recent experimental work by Kontos \textit{et al.} \cite{Kontos13-PRC}, an extensive R-matrix analysis was also performed using their new results as well as the \mbox{post-2000} capture data \cite{Singh04-PRL,Bemmerer06-PRL, Confortola07-PRC, Gyurky07-PRC, Costantini08-NPA,Brown07-PRC,DiLeva09-PRL, Carmona-Gallardo12-PRC, Bordeanu13-NPA}. Taking into account the internal capture contribution to the cross section, this R-matrix analysis successfully reproduced simultaneously the capture and scattering cross sections. The R-matrix study has been extended by deBoer \textit{et al.} \cite{deBoer14-PRC} with a detailed Monte Carlo analysis, which predicts the solar reaction rate with a 3-4\% precision; however, this is still not sufficient for, e.\,g., precise solar models. New experimental data are thus called for at higher energies.

Additionally, a recent study of the $^6$Li(p,$\gamma$)$^7$Be reaction showed a resonance like structure \cite{He13-PLB}. This has consequences on the level scheme of $^7$Be, where a positive parity level was proposed around $E_{x}=5.80$\,MeV excitation energy corresponding to $E_{cm}=4.21$\,MeV in the $^3$He($\alpha$,$\gamma$)$^7$Be reaction. This energy range was never investigated before, therefore there is no experimental constraint from this reaction on the proposed level.

In this paper we present a measurement of the $^3$He($\alpha$,$\gamma$)$^7$Be reaction cross section in the energy range of $E_{cm}=4.0 - 4.4$\,MeV with $\approx$ 0.04-MeV energy steps, to cover the range of the proposed resonance. The present experiment answers the call for a higher-energy dataset and also provides a constraint on the strength of a possible resonance in $^3$He($\alpha$,$\gamma$)$^7$Be corresponding to the proposed $^7$Be level.

The paper is organized as follows. In \Sec{sec:meth} details about the experiment are given. In  \Sec{sec:analy} the data analysis is described, in \Sec{sec:res} the experimental results are presented, and the impact of the new data on the properties of the proposed resonance is discussed in \Sec{sec:disc}. Finally a summary is given in \Sec{sec:sum}.

\section{\label{sec:meth} Experimental details}
For the cross-section determination, the activation technique was used.
The number of radioactive $^7$Be reaction products was determined via $\gamma$ spectroscopy by measuring the 477.6-keV $\gamma$ rays following 10.44\% of the $^7$Be decays \cite{Tilley02-NPA}.

\vspace{-3mm}
\subsection{Target} 
For the irradiations, a thin window gas cell was used, similar to the one in previous investigations \cite{Bordeanu13-NPA, Halasz16-PRC}. Both ends of the cell were closed by aluminum foils with 10-$\mu$m nominal thickness to confine the target gas (see \fig{fig:cell}). The entrance foil was pressed against an o-ring with a steel disk that had an opening of 12-mm diameter. The exit foil was glued on a steel disk with a hole of 14-mm diameter. A special heat-curing polymer was used for the glue\footnote{Vacseal{\textregistered} II Vacuum Leak Sealant}.

A specific entrance foil was typically used for a few subsequent irradiations. The closing foils served also as the catcher of the produced $^7$Be, therefore they were replaced after every irradiation. From the reaction kinematics the energy of the $^7$Be was calculated, and its energy loss was estimated using the SRIM code \cite{srim}. The implantation energy varied between $4.9-5.9$\,MeV, therefore the backscattering loss was negligible. The average range of $^7$Be in aluminum at these energies is between $6.9-8.3$\,$\mu$m, thus the catcher was thick enough to stop the activated particles. The $\alpha$ particles going through the cell were stopped in a water cooled tantalum sheet. 
The double window configuration was used to reduce the beam power deposition in surfaces with direct contact with the gas volume.

\vspace{-3mm}
\subsection{Irradiations} 
\begin{figure}[t]
\includegraphics[width=0.99\columnwidth]{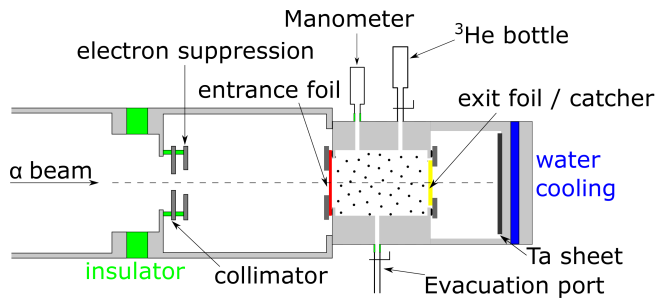}
\caption{\label{fig:cell} Schematic view of the activation chamber together with the thin window gas cell used for the experiment. (Not to scale.)}
\vspace{-2mm}
\end{figure}

For the irradiations, the MGC-20 cyclotron of Atomki was used. The gas cell was fixed onto an activation chamber acting as a Faraday cup. The chamber had a beam defining aperture with 5-mm opening and a secondary electron suppressing aperture supplied with $-300$\,V. The beam entered the chamber and gas cell through the apertures, which were isolated from the beam line allowing us to measure the beam current (see \fig{fig:cell}). The irradiation parameters are summarized in \tab{tab:param}, where the measured average electric current of the $\alpha^{++}$ beam is also shown.
\begin{figure}[t]
	\includegraphics[width=0.95\columnwidth]{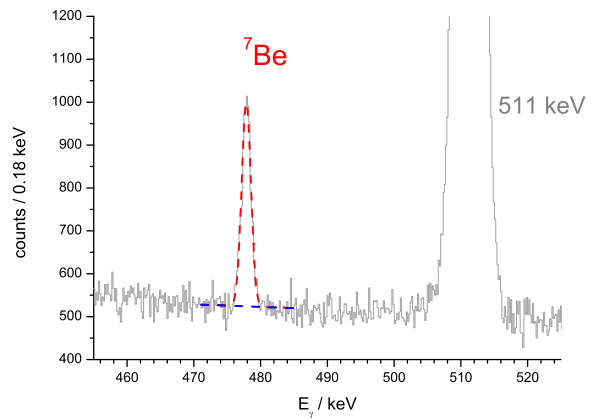}
\caption{\label{fig:spe} Spectrum recorded for 2.67 days after 20.95 days of waiting time following the 10.85-MeV irradiation. The fitted Gaussian peak and linear background are also shown. On the right side the partially visible annihilation peak is one order of magnitude greater than the small $^7$Be signal.}
\vspace{-2mm}
\end{figure}
\begin{figure}[!t]
\includegraphics[width=0.95\columnwidth]{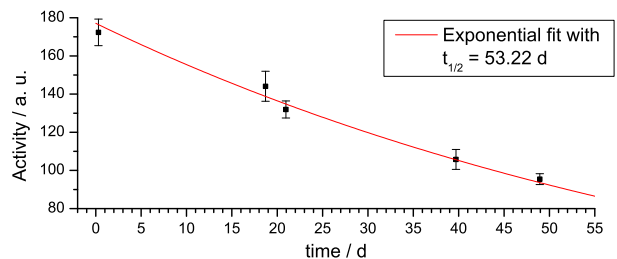}
\caption{\label{fig:decay} Measured activity based on the 477.6-keV peak area after the 10.85-MeV irradiation.}
\vspace{-2mm}
\end{figure}

\vspace{-3mm}
\subsection{\label{sec:counting} $\gamma$-ray countings}
After the irradiation the catcher foil implanted with $^7$Be was transported to a high-purity germanium detector with 100\% relative efficiency in commercial 10-cm-thick lead shielding. A single sample was counted for a few days, during which the area of the 477.6-keV gamma peak was detected with less than 5\% uncertainty. A typical spectrum is shown in \fig{fig:spe}. The counting was repeated multiple times, to follow the decay of the samples. No parasitic activity populating the 477.6-keV peak was observed; the decay curve shown in \fig{fig:decay} followed the half-life of $^7$Be (53.22\,days). The final cross-section result was derived from the sum of the spectra to reduce the statistical uncertainty.

\begin{table*}[!t]
\caption{Thickness of the entrance foil and target pressures, parameters of the irradiations including the calculated beam heating correction, and total lengths of the countings for a given data point.}
\label{tab:param}
\begin{ruledtabular}
\begin{tabular}{c | c c c | c c c | c}									
$E_{\alpha}$	&	Al thickness	&	Initial $^3$He pressure &	Final pressure	&	t$_{irr.}$	& I$_{avr.}$ 	 & BH correction 	 &  t$_{{count}_{tot}}$ \\
 / MeV			&	  / $\mu$m	&		/ mbars		 &		/ mbars &	/ h 			& / $\mu$A		 &	 / \%     &	/ days   					\\
\colrule								
7.30	&	10.80\,$\pm$\,0.10	&			101.6	&	102.3	&	20.00	&	0.51	&	0.7	&	14.78	\\							
10.45	&	10.62\,$\pm$\,0.09	&			105.5	&	116.6	&	23.50	&	0.69	&	0.7	&	\phantom{1}7.78	\\
10.60	&	10.81\,$\pm$\,0.04	&\phantom{1}98.9	&	109.8	&	15.80	&	0.72	&	0.7	&	13.84	\\
10.65	&	10.43\,$\pm$\,0.10	&			101.8	&	115.6	&	21.92	&	0.76	&	0.8	&	10.10	\\
10.75	&	10.50\,$\pm$\,0.03	&			101.3	&	117.6	&	22.33	&	1.27	&	1.3	&	\phantom{1}6.79	\\
10.85	&	10.27\,$\pm$\,0.09	&			101.5	&	112.6	&	18.00	&	0.93	&	0.9	&	17.07	\\
10.95	&	10.37\,$\pm$\,0.02	&			105.9	&	120.6	&	22.00	&	1.01	&	1.0	&	\phantom{1}6.58	\\
11.05	&	10.50,$\pm$\,0.03	&\phantom{1}99.4	&	118.8	&	22.03	&	1.27	&	1.3	&	\phantom{1}7.47	\\
11.15	&	10.37\,$\pm$\,0.02	&			100.6	&	115.6	&	22.00	&	1.43	&	1.4	&	\phantom{1}5.25	\\
11.25	&	10.50\,$\pm$\,0.03	&			101.2	&	119.9	&	19.15	&	1.22	&	1.2	&	\phantom{1}9.19	\\
11.35	&	10.37\,$\pm$\,0.02	&			101.3	&	105.9	&	18.30	&	1.25	&	1.1	&	\phantom{1}4.30	\\									
\end{tabular}									
\end{ruledtabular}									
\end{table*}

\section{\label{sec:analy} Data analysis}

\subsection{Target thickness} 

The areal number density of the target was determined from the measured initial pressure and temperature of the gas, and the known length of the cell. The target thickness was chosen to optimize reaction yield, while limiting the energy uncertainty.

About 100\,mbars 99.999\% isotopically pure $^3$He gas were filled in the cell before each irradiation (see \tab{tab:param}).
The pressure in the cell was continuously monitored with an MKS Baratron (modell 722B) capacitive manometer gauge with a precision of about 0.1\,torr at this pressure range. Without beam the pressure was measured to be constant for several days. During the irradiations an increase of not more than about 20\,mbars was observed with a saturating trend caused by air desorption of the stainless steel wall and Al foils during the irradiation. The increased pressure was lower when a previous irradiation was performed with the same entrance foil, and the cell was not exposed to air for a longer time. After switching off the beam the pressure stayed constant at the increased level. Since the cell was closed after the filling, the number of $^3$He atoms was considered to be constant.
The additional pressure is considered to be due to air, and was taken into account in the effective energy calculation (see \Sec{sec:energy}).
The local-density reduction of the gas, the so-called beam heating effect, is at maximum about $0.7-1.4$\% calculated with the equation given in \cite{Marta06-NIMA}. In \tab{tab:param}, the beam-heating correction factors for each run are given, as a conservative estimate the correction factor is also considered as statistical uncertainty in the target thickness determination (see \tab{tab:uncertainty}).

The temperature of the cell was constant at 23\,$\pm$\,2\,$^\circ$C determined by the cooling water flowing through the body of the cell.

The length of the cell between the foils with compressed o-rings is 41.5\,mm. The pressure difference between the beam line vacuum and target cell causes a deformation of the entrance and exit foils changing the length of the beam path in the cell and thus altering the number of target atoms at a given pressure. The deformation of the entrance and exit foils was measured off-site with a laser distance sensor as described in \cite{Halasz16-PRC}. The maximum deformation of both the entrance and exit foils was found to be less than 0.2 and 0.3\,mm, respectively. This total 0.5\,mm was added to the cell length for the effective gas length determination. As a conservative estimate, 1.0\,mm is taken as the uncertainty of the total target length stemming from the deformation and the uncertainty of the cell length measurement.

\subsection{\label{sec:energy}Effective energy}

\begin{figure}[t]
\includegraphics[width=0.95\columnwidth]{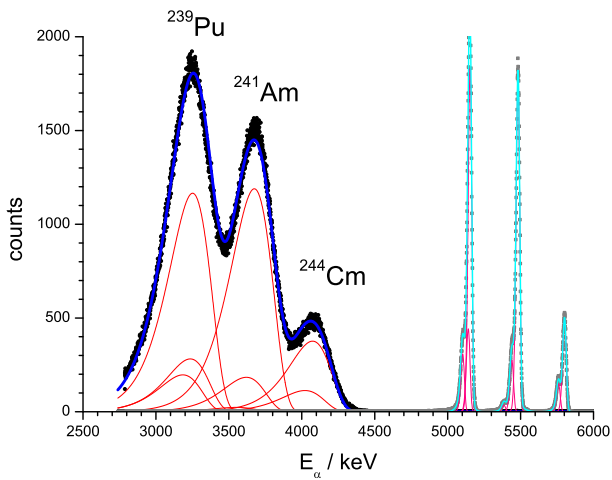}
\caption{\label{fig:Al} Example of a triple-nuclide source $\alpha$ spectra used for the Al thickness calculation. Black and gray points are the measured energy spectra with and without the aluminum foil, respectively. The energy calibration fit is plotted with light blue, while the fit at the $\chi^2_{red}$ minimum is shown as a dark blue curve which is the sum of the eight red lines for each transition. See text for details.}
\end{figure}

\begin{figure}[t]
\includegraphics[width=0.95\columnwidth]{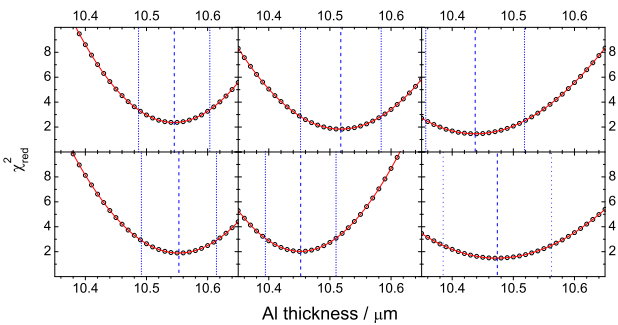}
\caption{\label{fig:chi} $\chi^2_{red}$ curves for a given entrance foil, measured at different spots. Vertical dashed and dotted lines show the minima and the 1\,$\sigma$ uncertainty, respectively. See text for details.}
\end{figure}

As the beam is decelerated in the entrance foil, the knowledge of the accurate window thickness is crucial in the reaction energy calculation.
The thickness of the entrance foils was measured by $\alpha$-energy loss, where the energy of the $\alpha$ particles from a triple-nuclide source (i.\,e., containing three $\alpha$ emitters as $^{239}$Pu, $^{241}$Am, and $^{244}$Cm) was measured after penetrating through the foils in a SOLOIST $\alpha$ spectrometer. As a first step, the $\alpha$ spectrum without the foil was measured and the eight peaks from this source were fitted with Gaussian plus low energy tail functions simultaneously for energy calibration (see gray points in \fig{fig:Al}). The peak area ratio for a given isotope was fixed at its literature branching ratio value. Then the spectrum of the $\alpha$ particles penetrating the foil was recorded and fitted as follows:
The energy of the $\alpha$ particles was calculated using SRIM \cite{srim} by assuming a given aluminum thickness. The center of the Gaussian peaks was fixed at each given energy, and as in the calibration case the branching ratios were also fixed. Only the total intensities of the peaks and the widths were varied in the fit. This kind of fit was repeated for many assumed Al thicknesses, and the reduced chi square ($\chi^2_{red}$) of each fit was calculated. The aluminum thickness-dependent $\chi^2_{red}$ values can be fitted by a parabola (see \fig{fig:chi}). The minimum of the fitted parabola was taken as the actual foil thickness. The thickness uncertainty was also calculated from the $\chi^2_{red}$ curve by taking the $min\,\chi^2_{red} + 1$ values\footnote{In case of  4085 degree of freedom (4096 data point and 11 parameters) the 1\,$\sigma$ uncertainty border is at $min\,\chi^2 + 4127$. These numbers were divided by the degrees of freedom.}.
The thickness of each entrance foil is measured on a few different points and consistent values are obtained. Assuming uniform thicknesses, the statistically weighted average of those measurements is shown in \tab{tab:param} for a given entrance foil. The foil thickness uncertainty includes the fit uncertainty after the averaging only and does not include the uncertainty in the stopping power. The latter is taken into account only in the beam energy calculation, thus avoiding double counting.

The effective energy and its uncertainty have been determined as follows.
The beam energy loss was calculated using the SRIM tables. To avoid double counting of the uncertainties, the uncertainty in the stopping power of 3.5\% according to SRIM was assigned only to the energy-loss value.
The energy loss in the gas volume was also calculated and a stopping power uncertainty of 4.4\% was assigned. The effective energy was taken at the middle of the target and its uncertainty is based on the cyclotron beam energy uncertainty of 0.3\% and the previously mentioned foil thickness and stopping power uncertainties. 
As a conservative estimate, the energy loss caused by the extra gas in the cell (about 0.5\,keV) was added to the uncertainty of the beam energy, even though it is negligible compared to effects on the beam energy due to the uncertain stopping power of the Al entrance foil (about 40-50\,keV).
In total, the energy uncertainty is about 0.5\% (see \tab{tab:uncertainty}).

\subsection{\label{sec:eff} $\gamma$-detection efficiency}

The detector efficiency calibration was performed with multi-line calibration $\gamma$ sources ($^{133}$Ba and $^{152}$Eu) of precisely known activity (0.75 and 1.00\%) at 10- and 27-cm sample to end-cap distance to avoid the true coincidence summing effect. The obtained efficiency points were interpolated with a log-log linear function to get the efficiency at the $^7$Be $\gamma$-peak energy of 477.6\,keV. The $\gamma$ attenuation in the Al catcher at this energy is negligible considering the typical implantation depths.
The efficiency at the actual counting distance of 1\,cm was determined with a stronger $^7$Be source, which was counted in each geometry. Taking into account the time elapsed between the countings and the known half-life of $^7$Be, efficiency scaling factors were derived. Scaling both from 10 and 27\,cm gave consistent results.
The uncertainty of the 1-cm efficiency finally used is the quadratic sum of the interpolation (1.5\%), the calibration source activity (1.0\%), and scaling factor (0.9\%) uncertainties.

\section{\label{sec:res} Results}

The reaction cross sections obtained at the given center of mass energies are summarized in \tab{tab:res}.
In the table the statistical uncertainty is given for each point as well as the systematic uncertainty. The components of the uncertainties are summarized in \tab{tab:uncertainty}.

One data point was measured at $E_{c.m.} = 2479$\,keV to compare the results with literature data and validate the setup. At this particular energy the new data point agrees with the data from previous investigations.

\begin{table}[htb]
\caption{The measured $^3$He($\alpha$,$\gamma$)$^7$Be reaction cross sections.}
\label{tab:res}
\begin{ruledtabular}
\begin{tabular}{c c c}									
$E_{\alpha}$ 		&	$E_{cm}^{eff.}$\,$\pm$\,$\Delta$$E_{cm}^{eff.}$	&	$\sigma$\,$\pm$\,$\Delta$$\sigma_{stat}$\,$\pm$\,$\Delta$$\sigma_{syst}$ 	\\
 / MeV				&	/ keV			&	/ $\mu$barn   \\
\colrule
\phantom{1}7.30	&	2479\,$\pm$\,23	&	\phantom{1}6.23\,$\pm$\,0.16\,$\pm$\,0.27	\\
10.45	&	3995\,$\pm$\,20	&	10.47\,$\pm$\,0.24\,$\pm$\,0.45	\\
10.60	&	4057\,$\pm$\,21	&	10.13\,$\pm$\,0.25\,$\pm$\,0.44	\\
10.65	&	4097\,$\pm$\,20	&	10.75\,$\pm$\,0.23\,$\pm$\,0.46	\\
10.75	&	4140\,$\pm$\,20	&	10.39\,$\pm$\,0.23\,$\pm$\,0.45	\\
10.85	&	4197\,$\pm$\,20	&	10.49\,$\pm$\,0.25\,$\pm$\,0.45	\\
10.95	&	4238\,$\pm$\,20	&	10.71\,$\pm$\,0.22\,$\pm$\,0.46	\\
11.05	&	4278\,$\pm$\,20	&	10.46\,$\pm$\,0.24\,$\pm$\,0.45	\\
11.15	&	4330\,$\pm$\,20	&	10.80\,$\pm$\,0.26\,$\pm$\,0.46	\\
11.25	&	4370\,$\pm$\,20	&	10.56\,$\pm$\,0.24\,$\pm$\,0.45	\\
11.35	&	4422\,$\pm$\,20	&	10.60\,$\pm$\,0.28\,$\pm$\,0.46	\\
\end{tabular}
\end{ruledtabular}						
\end{table}
	
\begin{table}[htb]
\caption{Uncertainty budget}
\label{tab:uncertainty}
\center
\begin{tabular}{l l}																		
Counting statistics:		& 1.6-2.3\%  \\
Temperature:			& 0.7\%  \\
Pressure:				& 0.3\%  \\
Beam heating:			& 0.7-1.4\%  \\
\hline
\bf Statistical uncertainty:& 1.9-2.8\% \\
\\
Detection efficiency:		& 2.0\%  \\
Gamma branching:		& 0.4\% \\
Beam current:			& 3.0\%  \\
Cell size:				& 2.4\%  \\
\hline
\bf Systematic uncertainty:& 4.3\% \\
\\
Initial beam energy:			& 0.3\%  \\
Al foil thickness:			& 0.2-1.0\%  \\
Energy loss in Al:		& 0.34-0.41\%  \\
Energy loss in the gas:	& 0.02\%  \\
\hline
\bf Energy uncertainty:	& 0.46-0.52\% \\								
\end{tabular}																
\end{table}
	
\section{\label{sec:disc} Discussion}

\begin{figure*}[tb]
\includegraphics[width=0.99\linewidth]{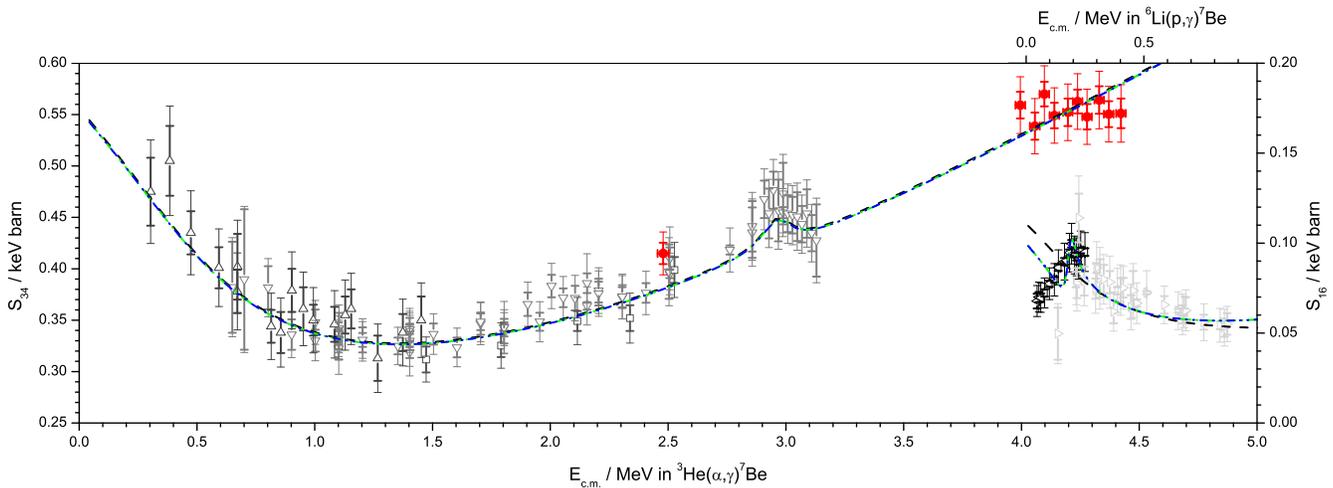}
\caption{\label{fig:fit} S-factor data from \cite{DiLeva09-PRL,Bordeanu13-NPA,Kontos13-PRC,Switkowski79-NPA,He13-PLB} shown as gray symbols and the present data shown as red dots. The thick error bars are statistical uncertainties used for the fits, while thin ones represent the total uncertainty of a given dataset. Black dashed, green dot-dashed and blue dot-dot-dashed lines are the constrained R-matrix fits assuming no extra level or $J_{\pi} = 1/2^+$ or a $J_{\pi} = 3/2^+$ level, respectively. On the left axis the scale of the S factor for the $^3$He($\alpha$,$\gamma$)$^7$Be reaction is shown, and on the right axis that for the $^6$Li($p$,$\gamma$)$^7$Be reaction is shown.  See text for details.}
\end{figure*}

As can be seen in \fig{fig:fit}, the new dataset does not show a peak-like structure in the energy range of \cite{He13-PLB}. Our result does not support the existence of a resonance in the $^3$He($\alpha$,$\gamma$)$^7$Be channel; however, within the statistical uncertainty of the dataset it may be present, but is not observable. Similarly no evidence of a level is visible in the scattering data covering this energy range, e.\,g., \cite{Tombrello63b-PR, Spiger67-PR}.
To quantify this limit, in the following the maximum allowed resonance strength is determined by a constrained R-matrix fit using a few datasets.

The R-matrix fit was performed with the AZURE2 code \cite{Azuma10-PRC}. Besides the data from this paper, three other $^3$He($\alpha$,$\gamma$)$^7$Be datasets from \cite{DiLeva09-PRL, Bordeanu13-NPA, Kontos13-PRC} were used. From the dataset of \cite{Kontos13-PRC} partial cross sections were used for the fit, but in the plot only the total cross sections are shown. Additionally, data from the $^6$Li($p$,$\gamma$)$^7$Be \cite{Switkowski79-NPA,He13-PLB} reaction channels were fitted simultaneously assuming either no level at $E_x=5.8$\,MeV or a $1/2^+$ ($3/2^+$) level as suggested in \cite{He13-PLB}. These datasets either include partial cross sections to the ground state and to the first excited state, or explicitly include the branching ratio between those for each data point. For the fit, the partial cross sections were used to better constrain the partial asymptotic normalization coefficients (ANCs), but in the plot only the total cross sections are shown.
As level parameters, those in \cite{Kontos13-PRC} were used. The positions and the widths of the levels including the background poles were kept fixed at their values presented there.
In the fit only the ANCs for both reactions and the $\alpha$, proton, and $\gamma$ widths of the proposed resonance (if included) were varied. The position of the resonance was fixed at $E_x=5.8$\,MeV.

The resulting fits are shown in \fig{fig:fit}. No apparent differences are seen for the $^3$He($\alpha$,$\gamma$)$^7$Be reaction cross section, and the fitted curves overlap. 
The fitted resonance parameters are shown in \tab{tab:fit}.

\begin{table}[tb]
\caption{ANCs and resonance parameters resulting from the R-matrix fit. The negative sign indicates negative interference. The calculated resonance strengths are also presented.}
\label{tab:fit}
\begin{ruledtabular}
\begin{tabular}{c | c c c}																		
$J_{\pi}$							&	w/o res.					&	 $1/2^+$			&	 $3/2^+$ \\
\colrule
ANC$_{p(1/2)}$(0) 					&	1.7\,fm$^{-1/2}$	&	2.1\,fm$^{-1/2}$	&	2.6\,fm$^{-1/2}$	\\
ANC$_{p(3/2)}$(0)					&	-2.2\,fm$^{-1/2}$	&	-2.6,fm$^{-1/2}$	&	-2.1\,fm$^{-1/2}$	\\
ANC$_\alpha$(0)					&	3.9\,fm$^{-1/2}$	&	3.9\,fm$^{-1/2}$	&	3.9\,fm$^{-1/2}$	\\
ANC$_{p(1/2)}$(429)				&	0.2\,fm$^{-1/2}$	&	1.8\,fm$^{-1/2}$	&	2.7\,fm$^{-1/2}$	\\
ANC$_{p(3/2)}$(429)				&	-2.9\,fm$^{-1/2}$	&	-2.7\,fm$^{-1/2}$	&	-1.8\,fm$^{-1/2}$	\\
ANC$_\alpha$(429)					&	3.0\,fm$^{-1/2}$	&	3.0\,fm$^{-1/2}$	&	3.0\,fm$^{-1/2}$	\\
$\Gamma_{p(1/2)}$					&	--					&	54\,keV			&	--		\\
$\Gamma_{p(3/2)}$					&	--					&	--					&	54\,keV	\\
$\Gamma_\gamma$(0)				&	--					&	-4.6\,meV			&	-2.3\,meV		\\
$\Gamma_\gamma$(429)			&	--					&	2.5\,meV			&	1.2\,meV		\\
$\Gamma_\alpha$					&	--					&	150\,meV			&	689\,meV	\\
\colrule			
calc. $(\omega\gamma)_\alpha$			&	--					&	 20\,neV				&	45\,neV	\\					
\end{tabular}									
\end{ruledtabular}
\end{table}

The resonance strength of the $^3$He($\alpha$,$\gamma$)$^7$Be reaction is calculated from the widths with the following equation.
\begin{equation}
(\omega\gamma)_\alpha = \frac{2J_x+1}{(2j_{^3He}+1)(2j_{\alpha}+1)} \frac{\Gamma_\alpha\Gamma_\gamma}{\Gamma_p+\Gamma_\alpha+\Gamma_\gamma}
\label{eq:strength}
\end{equation}

where $J_x$ is the spin of the excited state in $^7$Be, $j_{^3He}=1/2$ and $j_{\alpha}=0$ are the spins of the target and projectile; $\Gamma_\alpha$ and $\Gamma_p$ are the $\alpha$ and proton widths of the given level, respectively; and $\Gamma_\gamma=\Gamma_\gamma$(0)$+\Gamma_\gamma$(429) is the total $\gamma$ width.

The resulting $^3$He($\alpha$,$\gamma$)$^7$Be cross-section fits barely differ, and all of them are consistent with the present data, therefore the derived strengths can be treated as upper limits.

It is also important to mention, that the fitted S factor has an increasing slope in each fit, and does not become constant as the data suggest. With the given number of background poles (i.\,e., six at 11\,MeV and one at 7\,MeV) keeping their $\alpha$ widths around their Wigner limits and $\gamma$ widths below the Wiesskopf estimate, a constant S factor cannot be obtained. Further investigation of the R-matrix parameter space will be necessary to achieve a better fit, which is out of the scope of this paper.
Scattering data in a broader energy range could constrain even better the level widths, and additional capture data on either the $^3$He($\alpha$,$\gamma$)$^7$Be or $^6$Li($p$,$\gamma$)$^7$Be reactions would fix the $\gamma$ widths.

\section{\label{sec:sum}Summary}
The cross section of the $^3$He($\alpha$,$\gamma$)$^7$Be reaction has been measured around the proton separation energy of $^7$Be. The results show a constant cross section of about 10.5\,$\mu$barn in the region of the recently proposed  $^7$Be level.
By performing a constrained R-matrix fit, upper limits on the $\alpha$ capture resonance strength can be derived. If the resonance in the $^3$He($\alpha$,$\gamma$)$^7$Be reaction exists, its strength must be lower than $45$\,neV.

\begin{acknowledgments}
This work was supported by National Research, Development and Innovation Office (NKFIH) (Grants No. K120666 and No. NN128072), and New National Excellence Program of the Human Capacities of Hungary (Project No. \'UNKP-18-4-DE-449).
G.\,G.\,Kiss acknowledges support from the J\'anos Bolyai research fellowship of the Hungarian Academy of Sciences. The authors thank Dr. A.\,J. Timothy Jull for proofreading the paper.
\end{acknowledgments}
\bibliography{d:/Munka/Bibliografia/tszucs}
\end{document}